# Improving Collaborative Filtering based Recommenders using Topic Modelling

Jobin Wilson
R&D Department
Flytxt Mobile Solutions Pvt. Ltd.
Trivandrum-695581, India
jobin.wilson@flytxt.com

Santanu Chaudhury
Department of Electrical Engineering
Indian Institute of Technology, Delhi
Hauz Khas, New Delhi-110 016, India
santanuc@ee.iitd.ac.in

Brejesh Lall
Department of Electrical Engineering
Indian Institute of Technology, Delhi
Hauz Khas, New Delhi-110 016, India
brejesh.lall@ee.iitd.ac.in

Prateek Kapadia
R&D Department
Flytxt Mobile Solutions Pvt. Ltd.
Trivandrum-695581, India
prateek.kapadia@flytxt.com

## ABSTRACT
Standard Collaborative Filtering (CF) algorithms make use of interactions between users and items in the form of implicit or explicit ratings alone for generating recommendations. Similarity among users or items is calculated purely based on rating overlap in this case, without considering explicit properties of users or items involved, limiting their applicability in domains with very sparse rating spaces. In many domains such as movies, news or electronic commerce recommenders, considerable contextual data in text form describing item properties is available along with the rating data, which could be utilized to improve recommendation quality. In this paper, we propose a novel approach to improve standard CF based recommenders by utilizing latent Dirichlet allocation (LDA) to learn latent properties of items, expressed in terms of topic proportions, derived from their textual description. We infer user's topic preferences or persona in the same latent space, based on her historical ratings. While computing similarity between users, we make use of a combined similarity measure involving rating overlap as well as similarity in the latent topic space. This approach alleviates sparsity problem as it allows calculation of similarity between users even if they have not rated any items in common. Our experiments on multiple public datasets indicate that the proposed hybrid approach significantly outperforms standard user Based and item Based CF recommenders in terms of classification accuracy metrics such as precision, recall and f-measure.

## Categories and Subject Descriptors
H.3.3 [**Information Search and Retrieval**]: Information Retrieval and Search – *information filtering*

## 1. INTRODUCTION
With the availability of increasingly large quantities of online digital information, users find it challenging to locate information that is relevant and exciting to them. Recommender systems provide an effective way for information filtering by utilizing historical preferences expressed by users, to discover useful information. Collaborative Filtering (CF) is widely used for generating recommendations in many domains [4]. Standard CF purely relies on user-item interactions expressed in the form of implicit or explicit ratings, without considering features of users or items involved. One reason for the widespread adoption and success of Standard CF algorithms in practical recommender problems is this domain agnostic nature of the technique. The basic principle behind standard CF based recommenders is that users tend to like items that are either highly rated by other users having interests similar to them (User Based CF) [5] or items which are similar to other items that they themselves rated highly (Item Based CF) [6].Since similarity calculation is done purely based on rating overlap, standard CF based recommenders fail to consider latent properties of users or items which may be influencing user's rating decision on items [3].This results in poor quality recommendations in domains with very sparse ratings [1].

Even though model based CF techniques such as matrix factorization utilizes a latent factor approach by transforming both users and items into a single latent factor space, it still makes use of rating data alone to perform this transformation [2].

In many recommender domains such as movies, news or electronic-commerce, considerable contextual data in the form of unstructured text describing items being recommended is available. For example, movie plots and genre gets captured as plain text; item descriptions get captured as text, within product catalogs in electronic commerce domain. Utilizing such contextual data to transform users and items into single a latent factor space and leveraging rating information along with these latent factors to generate recommendations is a largely underexplored area.

In this paper, we propose a novel approach to improve standard CF based recommenders by making use of contextual data available in the form of text description of items. We learn latent features of users and items through topic modeling. Combining latent space based similarity with rating overlap-based similarity, we proposed a hybrid similarity score to refine the neighborhood formation, which helps in alleviating sparsity problem as it allows calculation of similarity between users even if they do not have any overlapping ratings. Our experiments on Movielens 1M dataset and a random subset of Netflix dataset with 2 million ratings indicate that our proposed hybrid recommender approach produces significantly higher quality recommendations in terms of precision, recall and f-measure, when compared with standard User Based and Item Based CF.

Rest of the paper is organized as follows. Section 2 describes related work in this area. In section 3 we outline the proposed hybrid recommender approach. Section 4 explains experiment setup and results. We conclude the paper in section 5.

## 2. RELATED WORK

Recommender systems are broadly classified into three categories: Collaborative Filtering (CF) Based, Content Based and hybrid [4]. CF based recommender systems make use of user-item interactions in the form of ratings as the basis for generating recommendations.CF is further classified into neighborhood based and latent factor based models [2].Neighborhood based CF tries to detect similarity between users or items based on the rating overlap. In a User Based CF, ratings given by similar people on an items is used to estimate a user's preference for those items [5] where Item Based CF exploit similarity of items with other items that the user has already rated to predict the user's preference on items [6]. Latent factor based CF models make use of Singular Value Decomposition (SVD) to factorize user-item rating matrix to determine latent properties of users and items [2].

Content Based recommender systems estimate user preferences against each of the content features from historical user ratings and the item properties, to construct a user profile [7]. New items are recommended to users based on the similarity of the item's content features with the constructed user profile, representing the user preferences. Hybrid recommender systems combine multiple approaches to improve quality of recommendations [4].

Latent Dirichlet allocation (LDA) is an unsupervised probabilistic generative model for modeling large text corpus [8]. It models each document as a mixture of topics which are latent and each topic as a mixture of words. Hybrid recommender systems have been proposed which make use of latent factor models based on LDA. Chang, Te-Min, and Wen-Feng Hsiao (2013) proposed an LDA based document recommendation system which utilized an Item Based CF algorithm with document similarity calculation based on latent topic distribution of documents [1]. Liu, Qi, et al (2012) proposed a latent factor model based on LDA to model evolution of user interests based on personalized ranking [3].

Our approach utilizes LDA to infer latent properties of items from their textual descriptions and then calculates user's preferences or persona in the same latent topic space based on historical ratings. We compute a hybrid user similarity score, which make use of user similarity in the latent topic space along with user similarity based on rating overlap to refine the user neighborhood. This way, our approach differs from the above references by simultaneously using user persona constructed using latent factors from item text descriptions as well as rating overlap based similarity of users to estimate a better user neighborhood to improve quality of recommendations.

## 3. PROPOSED APPROACH

Objective of this research is to extend standard CF algorithms to utilize textual description of items being recommended which is available in many domains, along with rating overlap to alleviate sparsity problem and improve quality of generated recommendations. Each item to be recommended is represented as a document containing a textual description about that item. We perform LDA on a corpus of such item documents to discover document-topic probability distribution as well as topic-word probability distribution. We represent each item in this latent topic space using the document-topic distribution as its feature vector. We add up item-topic distributions multiplied by normalized user rating, corresponding to each user's interests, to generate each user's topic-distribution vector, which indicates his persona in the same latent topic space. Once all users are represented in the latent topic space, similarity between users is calculated as a product of rating overlap based similarity and latent topic similarity. While building user neighborhood, we make use of his hybrid similarity metric as opposed to the standard rating overlap based similarity.

### 3.1 Discovering User Persona

Once LDA is performed over the corpus of item documents, discovering user persona in the same latent topic space is straight forward as described below.

**Step1:** Load all the $I$ item-topic distribution vectors into memory

**Step2:** For each user $U$, lookup & load the list of items that he has expressed interest on, into a list $L$

1. Initialize the current user $U$'s topic-distribution vector to zeros.
2. For each item $i$ in $L$ (each item he expressed interest on),
   1. Add the topic distribution vector for $i$, multiplied by $U$'s rating normalized by sum of all ratings from $U$, into $U$'s topic-distribution vector

### 3.2 Hybrid User Neighborhood Based Recommender

Once user persona is discovered in the latent topic space, we can easily find similarity between any pair of users in the latent topic space from symmetric Kullback–Leibler divergence between there latent topic distributions.

$$latent\_topic\_similarity(U, U') = e^{-KL(U_f, U'_f)}$$

$$KL(U_f, U'_f) = KL(U_f || U'_f) + KL(U'_f || U_f)$$

$$KL(U_f || U'_f) = \sum_i ln\left(\frac{U_f(i)}{U'_f(i)}\right) U_f(i)$$

Here $U_f$ denotes topic distribution indicating user $U$'s discovered persona. Symmetric KL divergence between latent topic distributions of users is translated into a similarity score between them using an exponential function as indicated above. This transformation ensures that we get a similarity value within the interval [0, 1].

Rating overlap based similarity among users is calculated using the standard approaches used in CF algorithms (by calculating Pearson correlation coefficient or Log-likelihood). Recommendations are generated using our proposed hybrid user neighborhood based approach as described below.

### 3.2.1 Hybrid Recommender Algorithm
For each user $U$, Perform the following steps

**Step1:** Build a neighborhood of size $N$ consisting of the most similar users to the current user, where similarity is defined as

$$sim(U, U') = topic\_sim(U, U') * rating\_overlap\_sim(U, U')$$

**Step2:** Generate a candidate list $L$ of all distinct items in the system that at least one of the user in the current user's neighborhood has expressed an interest on.

1. For each item $i$ in $L$
    1. Generate the *total_weight* for that item as
       tot*al_weight* =
       *(total_number_of_people_who_liked_it_in_neighborhood /neighborhood_size)*
2. Sort $L$ in descending order on the basis of total_weight.
3. Filter out all the items in $L$ which the current user has already expressed an interest on.
4. Pick the top $K$ items & return as recommendations

## 4. EXPERIMENTS & RESULTS
In this section, we compare performance of our proposed hybrid recommender approach with standard User Based CF and Item Based CF on two standard benchmark datasets: Movielens 1M dataset and Netflix dataset. Comparison is performed in terms of standard classification accuracy metrics: precision, recall and f-measure. Since we rely on textual description of movies in terms of plot and genre information, we make use of alternate sources such as IMDB interfaces [9] and OMDBAPI [10] for retrieving movie plots and genres.

### 4.1 Dataset Preparation
In this section, we describe various steps involved in preparing Movielens 1M [11] & Netflix datasets by using pre-processing and consolidation steps, for our experiments.

#### 4.1.1 Movielens 1M Dataset
We make use of Movielens 1M dataset along with IMDB dataset from IMDB interfaces in our experiments. We take the rating data from ratings.dat file and movie data in terms of title and genre from movies.dat file within Movielens 1M dataset. To obtain plots corresponding to all the movie titles in our dataset, we parse Plot.list file from IMDB dataset and use movie title and year to locate movie plots. In many cases, we observed that titles were not matching exactly due to which, data consolidation process required parsing along with human verification in case of ambiguous matches. We generated text files containing plot and genre information corresponding to each movie in our dataset. We also generated a random split of rating data into a training split with 80% of ratings from each user and a testing split with remaining 20% of each user's ratings. A total of 6040 users and 3706 movies exist in the rating dataset. Out of 3706 movies, 3266 titles have their plots available in IMDB Plot.list file.

#### 4.1.2 Netflix Dataset
We make use of Netflix dataset along with plot and genre information from OMDBAPI for our experiments. We take movie title and year of release from movie_titles.txt file within Netflix dataset and use OMDBAPI to query for plot and genre information. Out of 17770 movies from Netflix dataset, details of 13768 movies were available with OMDBAPI. A random sample of rating data with more than 2 million ratings were extracted from the Netflix dataset for our experiments. This dataset contains rating data pertaining to 5000 users and 17770 movies. We constructed text files containing plot and genre details corresponding to each of the 13768 movie titles to form the item document corpus corresponding to Netflix dataset. We also generated a random split of rating data into a training split with 80% of ratings from each user and a testing split with remaining 20% of each user's ratings.

### 4.2 Building Item Profiles
We fit an LDA model on the item document corpus corresponding to each dataset separately, to extract the corresponding item-topic probability distribution and word-topic probability distribution. We make use of MALLET [12] to perform LDA. Number of topics are fixed as 50 and model hyper-parameters alpha and beta are set to their default values 50 and 0.01 respectively. Figure 2 depicts few top keywords which got associated with each topic, for Movielens 1M dataset. Document-Topic distribution corresponding to each movie can now be considered as a representation of the movie in a latent feature space with 50 dimensions where each topic probability is a feature, indicating how strongly that topic represents that movie.

Figure 1 depicts the item-topic distribution corresponding to the popular movie *Schindler's List* from Movielens 1M dataset. It is interesting to observe that even though topics are latent and doesn't have a direct real-world interpretation, top keywords corresponding to each topic allows us to have some form of an interpretation of the item profile discovered by the model.

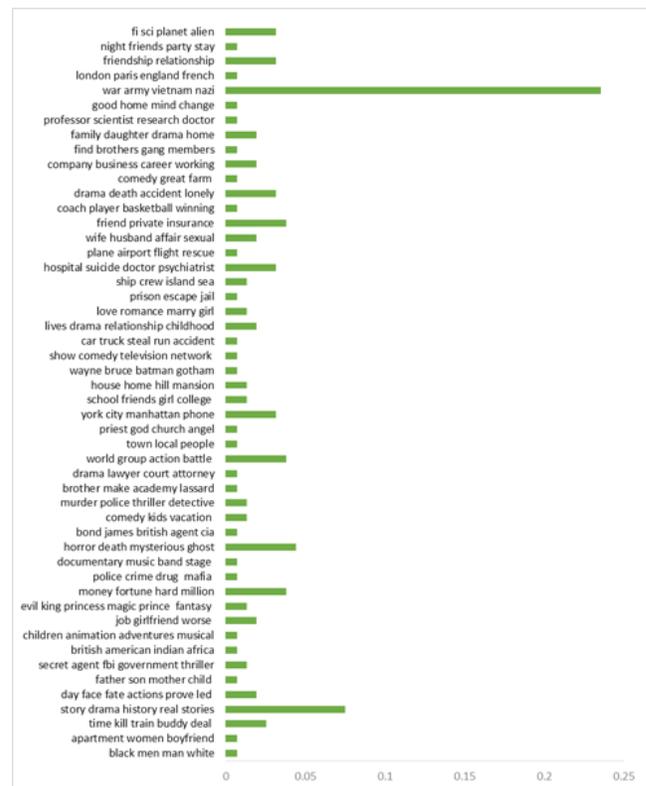

**Figure 1. Item-Topic Distribution for Schindler's List**

### 4.3 Building User Persona
We make use of the item profile discovered by LDA model to build user persona in the same latent topic space using the approach described in section 3.1.

| | |
|---|---|
| T0 | black men man white |
| T1 | apartment women boyfriend |
| T2 | time kill train buddy deal |
| T3 | story drama history real stories |
| T4 | day face fate actions prove led |
| T5 | father son mother child |
| T6 | secret agent fbi government thriller |
| T7 | british american indian africa |
| T8 | children animation adventures musical |
| T9 | job girlfriend worse |
| T10 | evil king princess magic prince fantasy |
| T11 | money fortune hard million |
| T12 | police crime drug mafia |
| T13 | documentary music band stage |
| T14 | horror death mysterious ghost |
| T15 | bond james british agent cia |
| T16 | comedy kids vacation |
| T17 | murder police thriller detective |
| T18 | brother make academy lassard |
| T19 | drama lawyer court attorney |
| T20 | world group action battle |
| T21 | town local people |
| T22 | priest god church angel |
| T23 | york city manhattan phone |
| T24 | school friends girl college |
| T25 | house home hill mansion |
| T26 | wayne bruce batman gotham |
| T27 | show comedy television network |
| T28 | car truck steal run accident |
| T29 | lives drama relationship childhood |
| T30 | love romance marry girl |
| T31 | prison escape jail |
| T32 | ship crew island sea |
| T33 | hospital suicide doctor psychiatrist |
| T34 | plane airport flight rescue |
| T35 | wife husband affair sexual |
| T36 | friend private insurance |
| T37 | coach player basketball winning |
| T38 | drama death accident lonely |
| T39 | comedy great farm |
| T40 | company business career working |
| T41 | find brothers gang members |
| T42 | family daughter drama home |
| T43 | professor scientist research doctor |
| T44 | good home mind change |
| T45 | war army vietnam nazi |
| T46 | london paris england french |
| T47 | friendship relationship |
| T48 | night friends party stay |
| T49 | fi sci planet alien |

**Figure 2. Topic-Word Distribution Sample: Movielens-IMDB**

Since user persona in the latent space is expressed as a probability distribution over latent topics, direct real-world interpretation of the profile may be difficult as in the case of item profile. However, top keywords corresponding to each topic allows a basic interpretation of the user profile. Figure 3 depicts the persona of user 5988 discovered by our model, from Movielens 1M dataset.

### 4.4 Generating Recommendations

To establish a baseline for our comparative study, we generate recommendations using standard User Based CF and Item Based CF using their implementations in Apache Mahout [13]. We use two different similarity measures namely Pearson Correlation Similarity and Log Likelihood Similarity in these standard CF experiments to study their impact on quality of recommendations. We also generate recommendations by using a variant of our proposed approach, in which we make use of user similarity in latent topic space alone in building user neighborhood. This allows us to study the impact of forming a hybrid neighborhood using rating overlap as well as similarity in latent topic space. For computing rating overlap based similarity between users, we again make use of Apache Mahout's implementation of Log Likelihood Similarity measure. Neighborhood size used in standard User Based CF as well as in our proposed approach is set as 30 in our experiments. We generate up to 75 recommendations per user, corresponding to each approach we described above, for comparing quality of recommendations.

### 4.5 Evaluating Recommendations

We use standard classification accuracy metrics namely precision, recall and f-measure as per their standard definitions in the recommender systems context [14]. We measure these values per user by retrieving $K$ recommendations (where K varies from 5 to 75) and calculate their average values across users, corresponding to each $K$. Recommendations are generated based on the training split of rating data and quality of prediction is assessed by comparing predictions with the test split of rating data. We compare trends corresponding to each of these metrics for different values of $K$, across different recommender approaches being evaluated.

### 4.6 Results

In this section, we present results from our experiments on Movielens 1M dataset and Netflix dataset that we prepared as described in section 4.1. To establish a baseline quality of recommendations on each dataset in our study, training split of corresponding rating data is provided to standard User Based CF and Item Based algorithms as described in section 4.4 to generate recommendations.

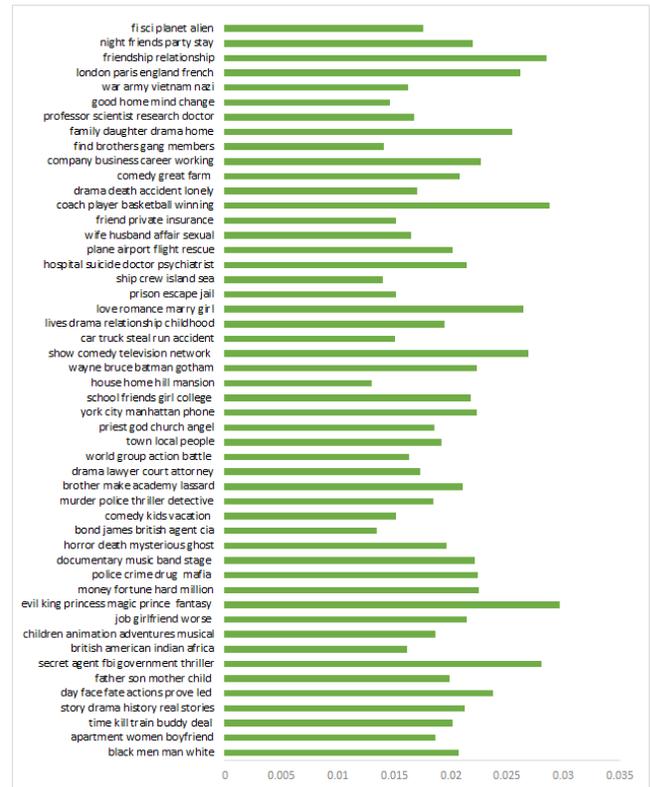

**Figure 2. User Persona: Movielens 1M User 5988**

Quality of predictions from each approach is evaluated as described in section 4.5. Table 1 below summarizes properties of training and testing splits of rating data corresponding to each dataset. Max. Ratings PU(maximum ratings per user) indicate maximum number of ratings that any user has given and Avg. Ratings PU(average ratings per user) indicate number of ratings that a user gives on an average.

**Table 1. Dataset Properties**

| Dataset | Users | Items | Max. Ratings PU | Avg. Ratings PU |
|---|---|---|---|---|
| Movielens (Training) | 6040 | 3677 | 1851 | 132.48 |
| Movielens (Testing) | 6040 | 3468 | 462 | 32.11 |
| Netflix (Training) | 5000 | 16666 | 4859 | 366.6 |
| Netflix (Testing) | 4994 | 13584 | 1214 | 90.78 |

Analysis of precision, recall and f-measure corresponding to recommendations generated from Movielens 1M dataset is visualized in Figure 3, Figure 4 and Figure 5 respectively. Legends in the visualizations are to be interpreted as described below.

UBCF(LL) represents standard User Based CF with Log Likelihood as the similarity measure. UBCF(P) represents standard User Based CF with Pearson Correlation as the similarity measure. IBCF(LL) represents standard Item Based CF with Log Likelihood as the similarity measure. IBCF(P) represents standard Item Based CF with Pearson Correlation as the similarity measure. "Proposed" represent our novel hybrid recommender which make use of user similarity in the latent topic space along with rating overlap based similarity, to refine the neighborhood formation. "Proposed_variant" represent a variant of our hybrid recommender which make use of user similarity in the latent topic space alone to refine the neighborhood formation.

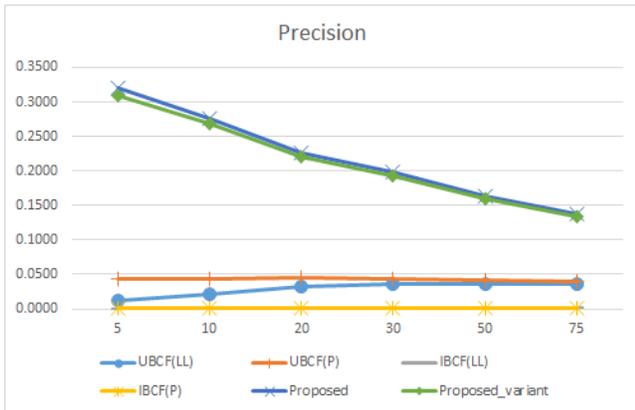

**Figure 3. Precision Analysis – Movielens 1M**

In Precision analysis of recommendations generated from Movielens 1M dataset, we observe that standard Item Based CF is performing the worst with precision values way less than even 1%. Standard User Based CF is generating values less than 5%. We observe that our proposed hybrid approach performs the best with precision value at 5 to be more than 31%. The variant of our hybrid approach which makes use of user similarity in latent topic space alone for neighborhood formation is also generating close by results, indicating that our approach of adding up item-topic distributions by scaling them proportional to the corresponding user rating for generating user persona is capturing some information on rating patterns also implicitly.

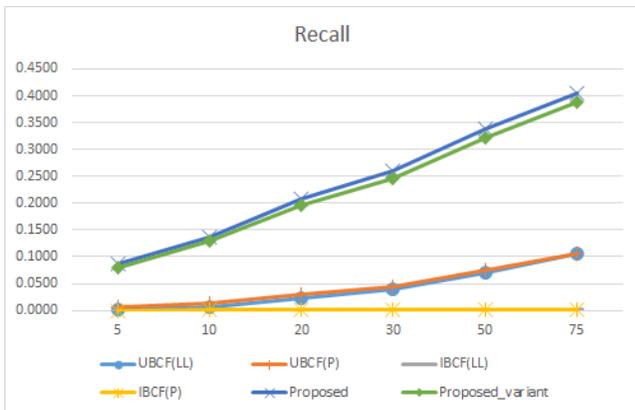

**Figure 4. Recall Analysis – Movielens 1M**

Recall analysis on recommendations generated from Movielens 1M dataset also indicate that our proposed hybrid recommender approach performs significantly better compared with other approaches. Recall at 30 indicate that our approach is able to retrieve more than 25% of relevant items where as standard User Based CF is able to only retrieve less than 5% of the relevant items.

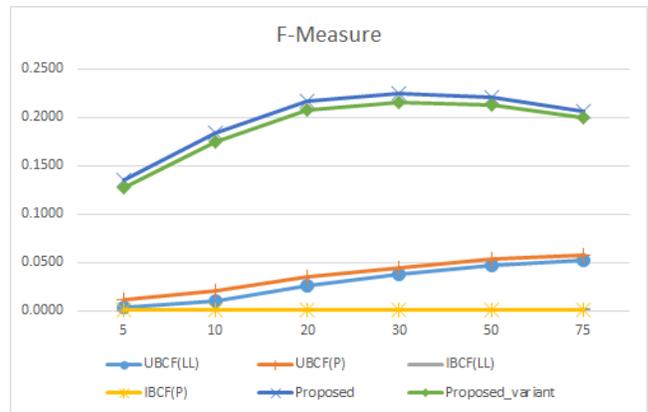

**Figure 5. F-Measure Analysis – Movielens 1M**

Analysis of F-Measure on recommendations generated from Movielens 1M dataset also indicates that our proposed hybrid approach is performing significantly better as compared to other approaches.

Analysis of precision, recall and f-measure corresponding to recommendations generated from Netflix dataset is visualized in Figure 6, Figure 7 and Figure 8 respectively.

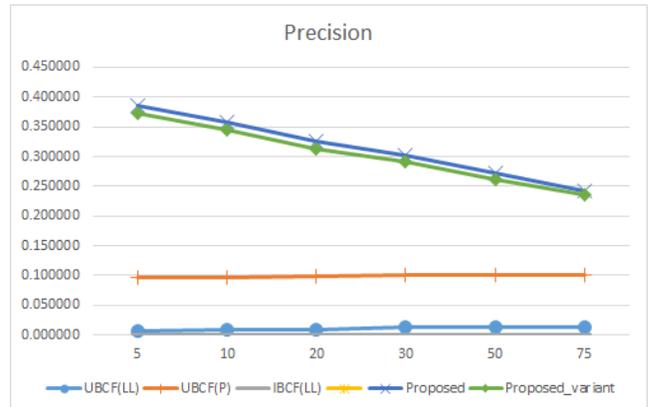

**Figure 6. Precision Analysis - Netflix**

Precision analysis of recommendations generated from Netflix dataset indicate that standard Item Based CF is performing the worst with precision values way less than even1%,as we observed in the case of Movielens 1M dataset. Standard User Based CF is generating precision values around 10%.

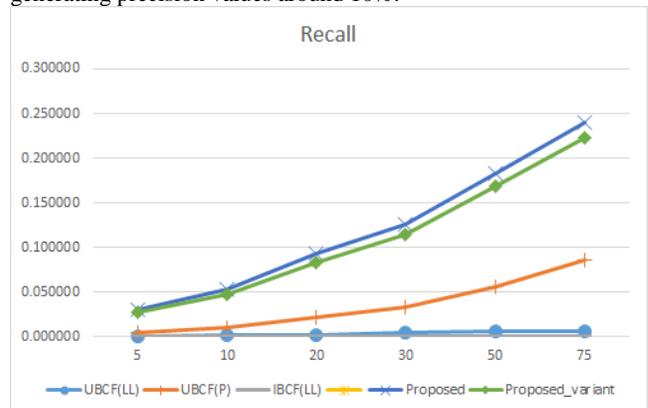

**Figure 7. Recall Analysis - Netflix**

Our proposed hybrid approach performs the best with precision value at 5 to be more than 38%. The variant of our hybrid approach which makes use of user similarity in latent topic space alone for neighborhood formation is also generating close by results as we observed in the case of Movielens 1M dataset.

Recall analysis on recommendations generated from Netflix dataset also indicate that our proposed approach performs significantly better compared to other approaches. Recall at 75 indicate that our approach is able to retrieve more than 24% of the relevant items where as standard User Based CF is able to retrieve only less than 9% of the relevant items.

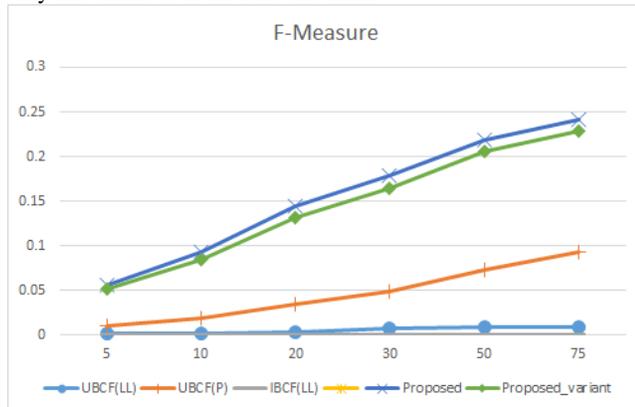

**Figure 8. F-Measure Analysis - Netflix**

F-measure analysis also ascertains that the proposed hybrid recommender approach is performing superior to other techniques, on Netflix dataset.

## 5. CONCLUSION
In this paper, we proposed a novel hybrid recommender approach using LDA, which utilizes similarity of users in a latent topic space along with their rating overlap based similarity to refine neighborhood formation, improving quality of recommendations. Our empirical evaluations indicate that the proposed approach significantly outperform standard User Based CF and Item Based CF and is well suited for recommender domains with contextual data in text form, describing items being recommended.